\documentclass[longauth]{aa} 
\usepackage{graphicx}
\usepackage{lscape}
\usepackage{longtable}
\usepackage{rotating}
\usepackage{natbib}

\usepackage{txfonts}
\newcommand{\mjup}{M$_{\rm Jup}$\,}

\newcommand{\msun}{M$_{\sun}$}

\newcommand{\bp}{$\beta$\,Pictoris\,}
\newcommand{\bpic}{$\beta$\,Pictoris\,}

\begin{document}
   \title{A narrow, edge-on disk resolved around HD~106906 with SPHERE
     \thanks{Based on data obtained with the VLT/Sphere at Paranal. Programme: 095.C-0298(A) }
    \thanks{This work is based (in part) on data products produced at the SPHERE Data Center hosted at OSUG/IPAG, Grenoble.}
}


   \author{  
     A.-M. Lagrange\inst{1,2} \and M. Langlois\inst{3,4} \and R. Gratton\inst{5} \and 
A.-L. Maire\inst{5,7} \and J. Milli\inst{6} \and J. Olofsson\inst{7,8,9} \and A. Vigan\inst{4,6} 
V. Bailey \inst{10} \and D. Mesa\inst{5} \and G. Chauvin \inst{1,2} \and  A. Boccaletti\inst{11} \and 
R. Galicher\inst{11} \and J.M. Girard\inst{6} \and M. Bonnefoy\inst{1,2} \and M. Samland\inst{7} \and 
F. Menard\inst{12} \and T. Henning\inst{7} \and M. Kenworthy\inst{13} \and C. Thalmann\inst{14} \and 
H. Beust\inst{1,2} \and J.-L. Beuzit\inst{1,2} \and W. Brandner\inst{7}  \and E. Buenzli\inst{14} \and 
A. Cheetham\inst{15} \and M. Janson\inst{16} \and H. le Coroller\inst{4,17} \and J. Lannier\inst{1,2} \and D. Mouillet\inst{1,2} 
\and S. Peretti\inst{15} \and C. Perrot\inst{11} \and G. Salter\inst{4} \and E. Sissa\inst{5} \and Z. Wahhaj\inst{6} \and L. Abe\inst{18} \and S. Desidera\inst{5} \and M. Feldt\inst{7} \and F. Madec\inst{4} \and 
D. Perret\inst{11} \and C. Petit\inst{19} \and P. Rabou\inst{1,2} \and C. Soenke\inst{7} 
\and L. Weber\inst{15}.   }

   \institute{
Univ. Grenoble Alpes, Institut de Plan\'etologie et d'Astrophysique de Grenoble (IPAG, UMR 5274), F-38000 Grenoble, France ; \email{lagrange@obs.ujf-grenoble.fr}
\and
CNRS, Institut de Plan\'etologie et d'Astrophysique de Grenoble (IPAG, UMR 5274), F-38000 Grenoble, France 
\and
CRAL, UMR 5574, CNRS, Universit\'e Lyon 1, 9 avenue Charles Andr\'e, 69561 Saint Genis Laval Cedex, France
\and
Aix Marseille Universit\'e, CNRS, LAM - Laboratoire d'Astrophysique de Marseille, UMR 7326, 13388, Marseille, France
\and
INAF -- Osservatorio Astronomico di Padova, Vicolo dell'Osservatorio 5, 35122 Padova, Italy
\and
European Southern Observatory, Alonso de Cordova 3107, Casilla 19001 Vitacura, Santiago 19, Chile
\and
Max Planck Institut f\"ur Astronomie, K\"onigstuhl 17, 69117 Heidelberg, Germany
\and
Instituto de F\'isica y Astronom\'ia, Facultad de Ciencias, Universidad de Valpara\'iso, Av. Gran Breta\~na 1111, Playa Ancha, Valpara\'iso, Chile
\and
ICM nucleus on protoplanetary disks, Universidad de Valpara\'iso, Av. Gran Breta\~na 1111, Valpara\'iso, Chile
\and
Steward Observatory, Department of Astronomy, University of Arizona, 933 North Cherry Avenue, Tucson, AZ 85721-0065, USA
\and
LESIA, Observatoire de Paris, CNRS, Universit\'e Paris Diderot, Universit\'e Pierre et Marie Curie, 5 place Jules Janssen, 92190 Meudon, France
\and
UMI-FCA, CNRS/INSU, France (UMI3386)
\and
Sterrewacht Leiden, P.O. Box 9513, Niels Bohrweg 2, 2300 RA Leiden, The Netherlands
\and
 Institute for Astronomy, ETH Zurich, Wolfgang-Pauli-Strasse 27, 8093 Zurich, Switzerland 
\and
Geneva Observatory, University of Geneva, Ch. des Maillettes 51, 1290, Versoix, Switzerland
\and  Department of Astronomy, Stockholm University, AlbaNova University Center, SE-106 91 Stockholm, Sweden
\and Observatoire de Haute-Provence, OH/CNRS, F-04870 St. Michel l'Observatoire, France
\and Laboratoire Lagrange, Université de Nice-Sophia Antipolis, Observatoire de la Côte d'Azur, CNRS UMR 7293, Nice Cedex 4, France
\and Onera - The French Aerospace Lab, 92322, Châtillon, France
}
 
   \date{Received date: this version XXX / Accepted date}

   
   \abstract
   {HD~106906AB is so far the only young binary system around which a planet has been imaged and a debris disk evidenced thanks to a strong IR excess. As such, it represents a unique opportunity to study the dynamics of young planetary systems.
}
   {We aim at further investigating the close  (tens of au scales) environment of the HD~106906AB system. 
   }
   {We used the  extreme AO fed, high contrast imager SPHERE recently installed on the VLT to observe HD~106906. Both the IRDIS imager and the Integral Field Spectrometer were used.
   }
   {We discovered a very inclined, ring-like disk at a distance of 65~au from the star. The disk shows a strong  brightness asymmetry with respect to its semi-major axis. It shows a smooth outer edge, compatible with ejection of small grains by the stellar radiation pressure. We show furthermore that the planet's  projected position is significantly above the disk's PA. Given the determined disk inclination, it is not excluded though that the planet could still orbit within the disk plane if at a large separation (2000--3000 au). We identified several additional point sources in the SPHERE/IRDIS field-of-view, that appear to be background objects. We compare this system with other debris disks sharing similarities, and we briefly discuss the present results in the framework of dynamical evolution.
}
{}
\authorrunning{A.-M. Lagrange et al}
\titlerunning{A disk resolved around HD~106906}

   \maketitle
%

\section{Introduction}
Circumbinary planets offer valuable constraints on planet formation theories \citep{thalmann14b}. Very few long period circumbinary planets are known today.  One of them is HD~106906, a Lower Centaurus Crux (LCC) member that hosts a massive (M=11$\pm$2 \mjup) giant planet (GP) detected in projected separation at 650 au by \cite{bailey14}. We recently demonstrated that HD~106906 is a close binary, therefore named HD~106906AB, with a total stellar mass probably greater than 2.5 \msun\, (Lagrange et al, 2015, subm.). In addition, a high-luminosity ($L_{d}/L_{*}$=1.4$\times$10$^{-3}$) circumbinary disk, indicated { by the near-infrared and far infrared SPITZER data,}  is also present  \cite[][]{chen05}, and the dynamical relation between the planet and the disk is therefore unknown. Given its youth \cite[13$\pm$ 2 Myr;][]{pecaut12}, this system offers unique opportunities to study early dynamics of planetary systems.  A very interesting related question is where and how the planet formed. If formed close to the star by core accretion  or disk gravitational instability, some mechanisms had to eject it on its current orbit, without destroying the disk material supposedly located at $\simeq$ 20 au. If formed bound to the star at its present location, an alternative formation mechanism should be invoked such as cloud collapse. Alternatively, the planet may have been captured from another star. This is a plausible scenario as the largescale (cluster) environment of HD~106906AB is and was certainly even denser at earlier ages (for a discussion, see Lagrange et al. 2015, subm.). Two related key questions are the position of the planet with respect to the disk, and the disk morphological properties.

As part of a large survey to search for planets around stars members of young and nearby associations, we recently recorded high contrast images of HD~106906AB with the SPHERE instrument recently mounted on the ESO's VLT Unit Telescope 3 \cite[][]{beuzit08}. The data resolve the disk for the first time, and  constrain the planet position relative to the disk. They also allow constraining precisely the GP population around the binary. This letter aims at presenting the observational results, and developing qualitative arguments on the system. We first describe the data and the observations (Section 2), then the results obtained on the disk (Section 3), the planet position relative to the disk (Section 4), and the search for additional planets in the system (Section 5). 

\section{The Data}
\label{}
%
\subsection{Data log}
Various images of HD~106906AB were recorded in March, May and July, 2015 with different instrumental set-ups (see Table~\ref{Logobs}). Exposures were taken in May at H band and in H2H3 dual band imaging \cite[respectively centered at 1.59 and 1.67 $\mu$m,][]{vigan10} with the IRDIS camera \cite[][]{dohlen08}, and at YJ (0.95-–1.35 $\mu$m, spectral resolution R $\simeq$ 54) with the Integral Field Spectrometer (IFS) \cite[][]{claudi08}. Note that these observations are unpolarized. In July, additional data were recorded in dual band imaging at K1K2 (centered at 2.11 and 2.25 $\mu$m) with IRDIS, and at YH (0.95-–1.65 $\mu$m, R $\simeq$ 33) with IFS. In all these observations, we used an apodized Lyot coronagraph including a 185 mas focal mask (SPHERE mask N${\_}$ALC${\_}$YJH${\_}$S) as well as a pupil mask. IRDIS provides a $\simeq$ 12\arcsec$\times$12\arcsec  field of view (FoV)~\cite[1100$\times$1100 au, given the star's distance 92.1 $^{+6.5}_{-5.7}$ pc;][]{vanleeuwen07}, with a $\simeq$ 12.2 mas/pixel scale. An IFS dataset consists of 21000 spectra spread over 5.1$\times$41 pixels on the detector. After extraction, the FoV is ~1.7\arcsec$\times$1.7\arcsec and the spaxel size is 7.4$\times$7.4 mas, i.e. 0.68$\times$ 0.68 au. 

The coronagraphic observations were performed keeping the pupil stabilized so as to perform Angular Differential Imaging (ADI) post-processing \cite[][]{marois06}. This allows the suppression of a large fraction of the residual starlight after the coronagraph. The FoV rotations for the different sets of data are provided in Table~\ref{Logobs}. Our observing sequence was as follows: 1/ Point Spread Function (PSF) imaging, with HD~106906AB offset from the mask, so as to record a PSF and a relative photometric calibration, 2/ Image of the star behind the mask, with four satellite footprints of the PSF that can be used for fine monitoring of the centering, 3/ Coronagraphic sequence, 4/ Image of the star behind the mask, with four satellite footprints again, 5/  PSF imaging, and 6/ Sky observations, with DITs corresponding to the DITs of the coronagraphic sequence. Finally, the True North (TN) and plate scales were measured using astrometric calibrators observed during each run, as part of the SPHERE GTO survey (Maire et al, 2015, subm.). They are reported in Table~\ref{Logobs}. Note that the plate scales and True North values were measured on non coronagraphic data. A shift of about 0.02 mas/pixel in plate scale has been found empirically between the non coronographic and coronagraphic data. For this reason, we will use conservative error bars for the separation measurements in the following. 

\begin{table*}
\caption{Observing Log. }
\label{Logobs}
\begin{center}
\renewcommand{\footnoterule}{}  
\small
\begin{tabular}{l l l l l l lll ll}
\hline 
Date& Set-up &DIT$\times$NDIT$\times$N
		&	Par. Ang. 
&Airmass
&	Seeing
&	Coh. time
	&	Wind
    & True North 
    & Plate scale
\\

& &(s) & ($\deg$)& & 	&  (ms) &(m/s)  & ($\deg$)&(mas/pixel)\\
\hline
2015/03/30&IRDIS$\_$H2H3 &64$\times$4$\times$9&-3.2/14.0 &1.17/1.18&$\geq$1.7	&1.2/1.4 & 14.5/18.3 & -1.8$\pm$ 0.1&  	12.255$\pm$ 0.008  \\
& IFS$\_$YJ& 64$\times$4$\times$9 & & &  &	&&  &  & \\
\hline
2015/05/06&IRDIS$\_$H& 16$\times$14$\times$16& -20.6/0.98 & 1.17/1.18&0.78/1.11& 1.7/2.2 & 4.1/9.6 &
-2.0$\pm$ 0.1& 12.220$\pm$ 0.003\\
& IFS$\_$YJ& 64$\times$4$\times$(7+9) & & &  &	&&  &  & \\
\hline
2015/05/12&IRDIS$\_$H2H3& 64$\times$4$\times$20  & -12.7/17.8&1.17/1.18 & 0.76/1.06 & 3.1/4.3	 & 2.2/5.7 &
-2.0$\pm$ 0.1& 12.220$\pm$ 0.03\\
& IFS$\_$YJ&64$\times$4$\times$16& & &  &	&&  \\
\hline
2015/07/03 &IRDIS$\_$K1K2 &64$\times$5$\times$16   &14.6/49.5&  1.18/1.29&0.64/1.36 &1.4/2.5&5.4/6.9	&   	
-1.8$\pm$ 0.15& 12.242$\pm$ 0.033\\
&IFS$\_$YJH& 64$\times$5$\times$16 & & &  &	  &  & \\

\hline
\end{tabular}
\end{center}
\tablefoot{{ DIT: Detector Integration Time. } Parallactic angle as measured at the start and end of each coronagraphic IRDIS sequence. Min and max values of the coherence time, airmass, windspeed, and seeing. Plate scales and True North were measured on non coronagraphic data.
}
\end{table*}

\subsection{Data reduction}
When observing with IRDIS in dual imaging at H2H3 (resp. K1K2), IRDIS produces two simultaneous images, the left one at H2 (resp. K1) and the right one at H3 (resp. K2). In classical imaging broadband H, two identical images are taken on the left and on the right detector quadrants. The IRDIS data were
corrected for cosmetics and sky background  using the SPHERE { Data Reduction and Handling (DRH)} pipeline \cite[][]{pavlov08}. The outputs include cubes of left and right images recentered onto a common origin using the satellite spots, and corrected from distorsion. After these first steps, the data were sorted out according to their quality, and the following algorithms were applied: { Classical Angular Differential Imaging \cite[cADI, ][]{marois06}, Template Locally
Optimized Combination of Images  \cite[TLOCI, ][]{marois14}), and Principal Component Analysis \cite[PCA, ][]{soummer12}). To do so, we used an analysis pipeline we developped to reduce and analyse our Guaranteed Time Observations }. The IFS data were also preprocessed using the DRH pipeline; then they were processed with PCA as described in \cite{mesa15}. For the May sequences, the data were median-binned in 16 (May, 12) and 14 (May 6) data cubes before reduction.

\section{A disk around HD~106906AB}

\begin{figure*}
\sidecaption
    \includegraphics[width=14.5cm]{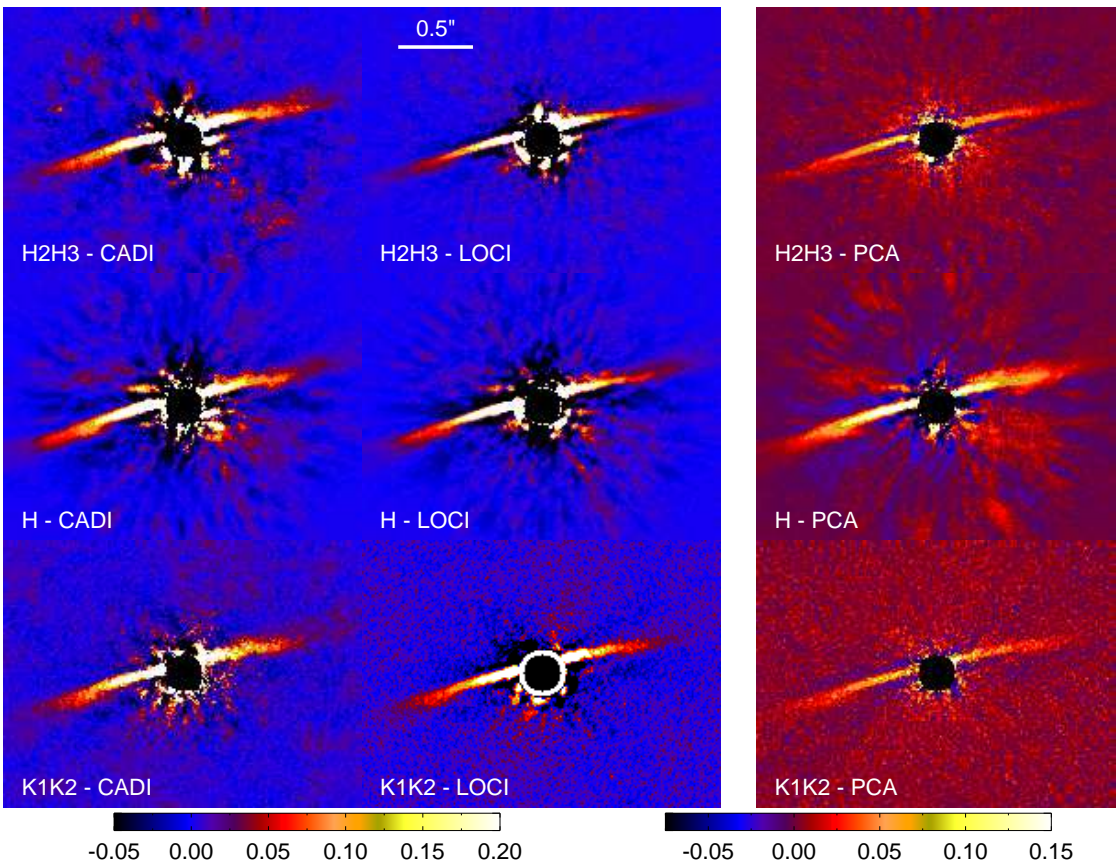}
 \caption{From left to right: CADI, LOCI, PCA, of the HD~106906AB disk at (from top to bottom) H2+H3, H and K1K2 bands (IRDIS data). North is up and East is to the left. The FoV of each image is 2.4\arcsec $\times$ 1.8\arcsec. The intensity-scale of the top, right image has been adapted to highlight the southern part of the disk.}
    \label{irdis_images}
  \end{figure*}

\begin{figure}[t!]
  \centering
\sidecaption
  \includegraphics[width=\hsize,angle=0]{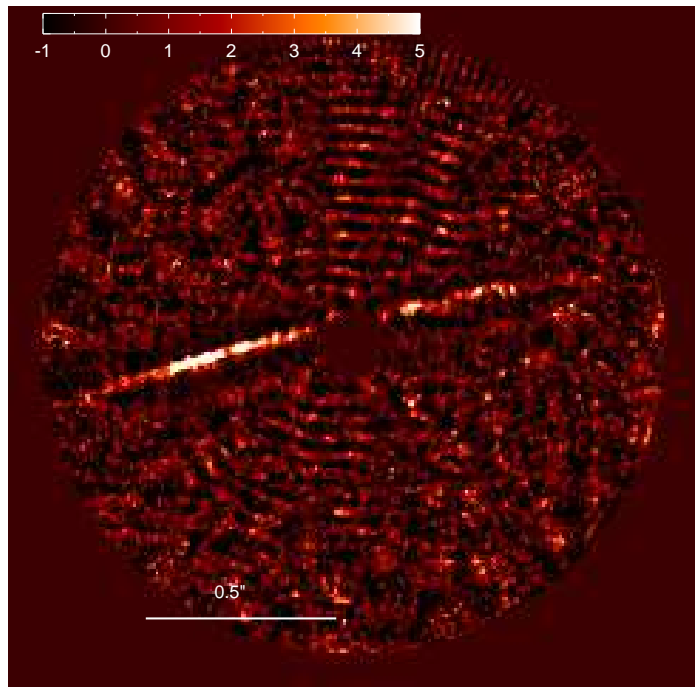}
    \caption{IFS YJH snr map of the HD~106906AB disk. N is up and E to the left.}
    \label{ifs_images}
  \end{figure}

All IRDIS data sets (even the March, 30th, of poor quality) and reductions reveal a close to edge-on disk. A set of resulting median images obtained using different algorithms is shown in Fig.\ref{irdis_images}. The IFS image resulting from the May, 12 data is shown in Fig.\ref{ifs_images}. The later image is a median over wavelength of the image obtained with PCA (with 50 components). A low-pass median filter (size of 31 pixels) and a software mask leaving the ring between 12 (0.09\arcsec) and 108 pixels (0.81\arcsec) from field center have been used. 

\subsection{Disk properties}
In our images, the disk appears mostly as a ring structure with a brightness distribution peaking at about 65 au, i.e., much further away than inferred from SED modeling under the assumption of pure blackbody grains \cite[20 au, ][]{bailey14}. It is seen out to 110 au in our images. It is elongated in the SE-NW direction. Its northern side is much brighter than the southern side, the southern side not being seen with all the reduction algorithms.  Given the age of the system as well as the absence of significant amounts of circumstellar gas, and its similarity with other debris disks (see below), we can safely conclude that the disk is a debris disk. 


A full modelling of the disk is beyond the scope of this Letter. We nonetheless performed a simple forward modelling of the disk images as done in \cite{milli12} using the GRATER code \cite[][]{augereau99} to constrain the disk morphology and disentangle ADI artifacts from real features. We modelled the disk as an inclined, optically thin ring, centered on the star, with a dust density distribution that peaks at a radius $r_0$ and follows a power law of slope $\alpha_{\mathrm{in}}$ within $r_0$ and $\alpha_{\mathrm{out}}$ beyond $r_0$. The model geometry is defined by six free parameters: the inclination $i$, the position angle (PA), the radius of peak dust density $r_0$, the Henyey-Greenstein coefficient g parametrizing the anisotropy of scattering, the outer slope $\alpha_{\mathrm{out}}$ and a scaling factor to match the disc total flux. We fixed the inner slope $\alpha_{\mathrm{in}} $ to 10 because our images are unable to constrain this parameter. Likewise, the vertical dust density distribution is set to a gaussian profile of scale height 0.5 au at $r_0$ with a linear flaring. 
For each set of data, the disk model is rotated to the appropriate parallactic angles, convolved with the instrumental PSF and subtracted from each frame. The resulting data cube is reduced using the same PCA algorithm as  described previously, retaining 8 components for the PSF subtraction. These steps are repeated by varying the free parameters until a reduced chi-squared is minimized. The reduced chi-squared is computed in an elliptical annulus where the disk is detected. The result of this minimization is given in Table \ref{disk_params}. We derive an inclination of 85.3 $\pm$ 0.1 deg, a disk position angle (PA) of 104.4 $\pm$ 0.3 deg, a Henyey-Greenstein parameter of anisotropy g of 0.6 $\pm$ 0.1. Note that we also performed a simple elliptical fit on the IRDIS data, which gave similar values for the inclination and PA.
The disk is strongly forward scattering. Finally, the inversion leads to a slope of -4 for the outer edge of the ring.

\begin{table}
\caption{Best model parameters for the HD~106906 disk after forward modeling.}
\label{disk_params}
\begin{center}
\small
\begin{tabular}{l l l l ll l l lll }
\hline 
Filter & $r_0$ (au) & incl. $i$ ($^\circ$)  & PA ($^\circ$) & g & $\alpha_\mathrm{out}$ \\
\hline
H2 &  66.0$\pm$1.8  & 85.4$\pm$0.1 & 104.4$\pm$0.3 & 0.6$\pm$0.1 &  -4.5$\pm$0.3 \\
H3 &  63.0$\pm$1.0  & 85.2$\pm$0.1 & 104.3$\pm$0.3 & 0.6$\pm$0.1 &  -3.8$\pm$0.4 \\
\hline
\end{tabular}
\end{center}
\end{table}

\begin{figure}[t!]
    \centering
   \includegraphics[width=\hsize,angle=0]{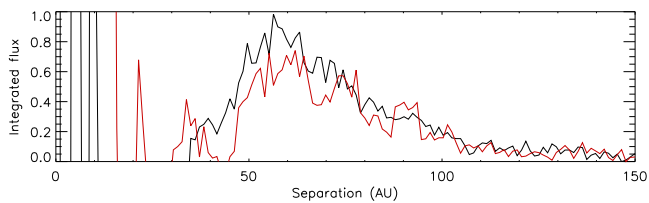}
    \caption{Disk brightness along the semi-major axis, extracted from H2+H3+K1+K2 data (see text). To improve the SN, we averaged the flux over 3 pixels perpendicular to this axis. The black curve corresponds to the East side, the red one to the West side. }
    \label{sbdsma}
  \end{figure}

A remarkable feature of the disk is its E-W brightness asymmetry, as clearly seen in Fig.\ref{irdis_images}, Fig.\ref{ifs_images} and as well in Fig.\ref{sbdsma}, where we show the brightness distribution along the South-East and North-West ansae as measured on the H2+H3+K1+K2 image. The SE side appears about 10$\%$ brighter than the NW side. Our simple symmetric model is not able  to fully account for this brightness asymmetry. This asymmetry could be explained by an elliptic disk, with a pericenter located on the NW side (but see also below). 

\subsection{Comparison with other debris disks}
The disk of HD~106906AB  shows similarities with the HR4796 one (\cite{schneider09}, \cite{lagrange12c}, see also Milli et al, 2015, in prep.) : they both present rings with an inner void, with similar distances and widths (Table\,\ref{compa_disks}). Yet, their outer edges are quite different: while the HR4796 disk brightness slope is about $-10$, the HD~106906AB disk is $-4$, i.e. more comparable to that of the \bpic disk \cite[][]{lagrange12a}. A full disk modeling of the \bpic system, including radiation pressure acting on the small grains produced by collisions, reproduced most of the disk asymmetries \cite[][]{augereau01}. In the case of HR4796, the steep edge is not fully understood; it may imply either the presence of a companion close to the outer disk  or be due to a higher opacity in the HR4796 disk that partly blocks the stellar flux \cite[for a detailed discussion, see][]{lagrange12c}.  

Neither the \bpic disk nor the HR4796 disk show strong side asymmetries as does the disk of HD~106906AB. The disk around the F2V star HD~15115 (see also Table\,\ref{compa_disks}) is more comparable from this point of view \cite[][]{mazoyer14}, with asymmetries both along the minor and major axes. Possible explanations to the asymmetry along the semi-major axis include interactions with the ISM, perturbation by a possible neighbour star, HIP~12545, or intrinsic disk properties. The HD~15115 disk is bowed \cite[see][]{rodigas12}; such a bow might be present in HD~106906 data, but the data are not good enough to allow a firm conclusion at this point. 

A comparison with the solar type star HD~61005 disk \cite[][]{hines07} is interesting as the disk also has a narrow ring-like shape, with a sharp inner edge. Brightness asymmetries are seen also along the semi-major axis and semi-minor axis. An extending, tenuous feature is seen emerging from the ring, possibly due to interactions with a warm, low density cloud \cite[][]{maness09}. Such extended features are not seen in our data.

The HD~106906 system is the second known system with a resolved disk and an imaged planet; the other is the \bpic system\footnote{Fomalhaut hosts a disk and a possible planet surrounded by dust, but the planet photons have not been detected yet \cite[][]{kalas08}}. In both cases, the planet is a massive giant and the disk bulk of material lies at a few tens of aus. In both cases, the inner part of the disk shows a relative void of material. The host stars are both early-type, massive, young stars. Yet, while HD~106906AB b is orbiting more than 600 au, \bpic b orbits at less than 10 au. The dynamical histories of these planets, and also their formation histories, might be different.

\begin{table*}
\caption{Morphological properties of the HD~106906, HR4796, HD15115, HD61005 and \bp disks. }
\label{compa_disks}
\begin{center}
\renewcommand{\footnoterule}{}  
\small
\begin{tabular}{l l l l ll l l lll }
\hline 
System & Disk luminosity & System age & Star ST & Star mass & Peak Sep.&FWHM  & $\alpha$$_{\mathrm{in}}$&$\alpha$$_{\mathrm{out}}$
&offset\\
&$L_{d}/L_{*}$& Myr&&(\msun) &(au)&(au)&&&\\
\hline
HD~106906AB&1.4$\times$10$^{-3}$&13$\pm$ 2 & F5V&2.6&65&30&10 (fixed)&-4 &?\\
HR4796&5$\times$10$^{-3}$& 8$\pm$ 2& A0V &2.4&80 & 10 &$\leq$ -3 & $\leq$-13 &  $\simeq$ YES\\
\bpic &3$\times$10$^{-3}$ & 21$\pm$ 3&A5V &1.75 &100 &$\geq$40 &$\simeq$-2 &-3.5 & NO\\
HD~15115 &5$\times$10$^{-4}$ & 12-100&F2V &1.6 &90 & &10 &-4 & YES\\
HD~61005 &2$\times$10$^{-3}$ & 90$\pm$ 40 & G8V& & 60& &$\simeq$5 &-4.5 & YES\\
\hline
\end{tabular}
\end{center}
\tablefoot{Ref. HD~106906: \cite{pecaut12}, this work. HR4796: \cite{lagrange12c}, \cite{milli15}, \cite{schneider09}, \cite{stauffer95}. \bpic: \cite{lagrange12b}. HD15115: \cite{kalas07}, \cite{mazoyer14}. HD65001: \cite{hines07}, \cite{buenzli10}, \cite{desidera11}.
}
\end{table*}


\section{Position of the planet relative to the disk}
\begin{figure}[t!]
    \centering
\includegraphics[width=\hsize,angle=0]{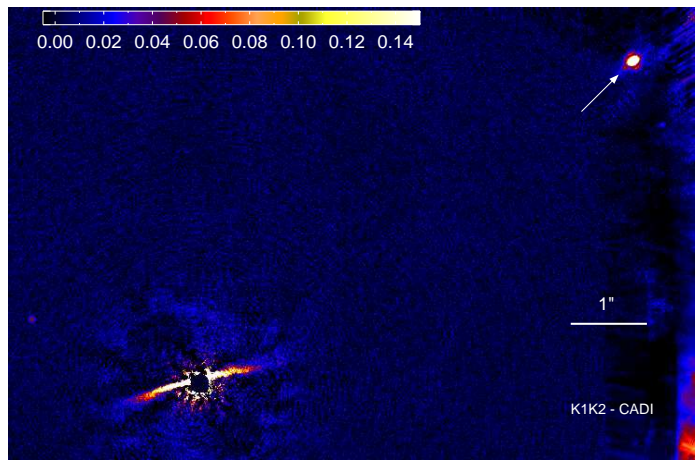}
    \caption{The disk and the planet around HD~106906AB. North is up and East is to the left. HD~106906AB b is the bright source top-right. }
    \label{K1K2_image}
  \end{figure}

HD~106906AB b is located more than 7" away from the central stars in projected separation \cite[][]{bailey14}. With such a separation, the planet is at the edge of our IRDIS FoV. We then carefully scheduled a series of observations to image both the disk and the planet on the same frames, to measure as precisely as possible the planet PA relative to the disk PA using a single data set. Indeed, when comparing PA coming from different instruments or telescopes, the systematics due to true North can be very high, as demonstrated in \cite{lagrange12a}.  We had then a timing constraint to ensure that the planet would be crossing the detector FoV in Pupil Tracking mode, and to ensure as well a large enough FoV rotation. The image resulting from this set of observations is  shown in Fig.\ref{K1K2_image}\footnote{In practice, the planet was within the FoV only in 35 cubes out of the whole data set (made of 64 cubes). We selected these 35 cubes and reduced them again. We checked that the position of the planet using the extracted dataset was the same as the one using the whole dataset.}. Using the NICI and HST ACS data, we confirm that HD~106906AB b shares a common proper motion with the central stars (see Fig.\ref{ppm} and Table \ref{cc}).
 
The first question addressed is whether the planet could still be orbiting within the disk, even though seen 23 degrees above the disk in projected position. Assuming a disk inclination of i$_{d}$, and a difference between the disk and the planet PA of i$_{p}$, then the planet could still be within the disk if located 650 $\times \sqrt{(1+ (\frac{sin(i_{p})}{tg(i_{d})})^{2})} $  au from the central stars. With i$_{d}$ = 5$^o$ (resp. 7$^o$), its physical separation to the central star would be $\simeq$ 3000 (resp. 2000) au. Then, the planet could still be within the disk plane if located at a much larger distance than the projected separation. We note that even if, on the contrary, the real inclination of HD~106906AB b is 23 degrees with respect to the disk, the planet would not induce Kozai resonance effects on the disk, whatever its orbital properties. Indeed, for most three-body systems, the Kozai mechanism starts only above a mutual inclination around $40\degr$ \cite[see][and ref. therein]{ford00}. Whether the planet may be causing, through regular perturbations, the observed disk asymmetry is not clear. To answer such a question, we would need to know the planet orbital properties, which will obviously be difficult to get. We conclude that even if complex, the stability of the binary + disk + planet system is probably not an issue. Indeed, the binary is very tight, so at the distance of the inner ring, it acts as a single, massive central star. Also, as seen above, if the orbit of HD~106906AB b is not coplanar with the disk, it should not destroy the disk provided its orbit is not too eccentric. To account for the inner edge of the disk, additional companions or other dynamic processes need to be invoked.

\section{Search for other planets}
\begin{figure*}[t!]
  \centering
\sidecaption
\includegraphics[width=12cm]{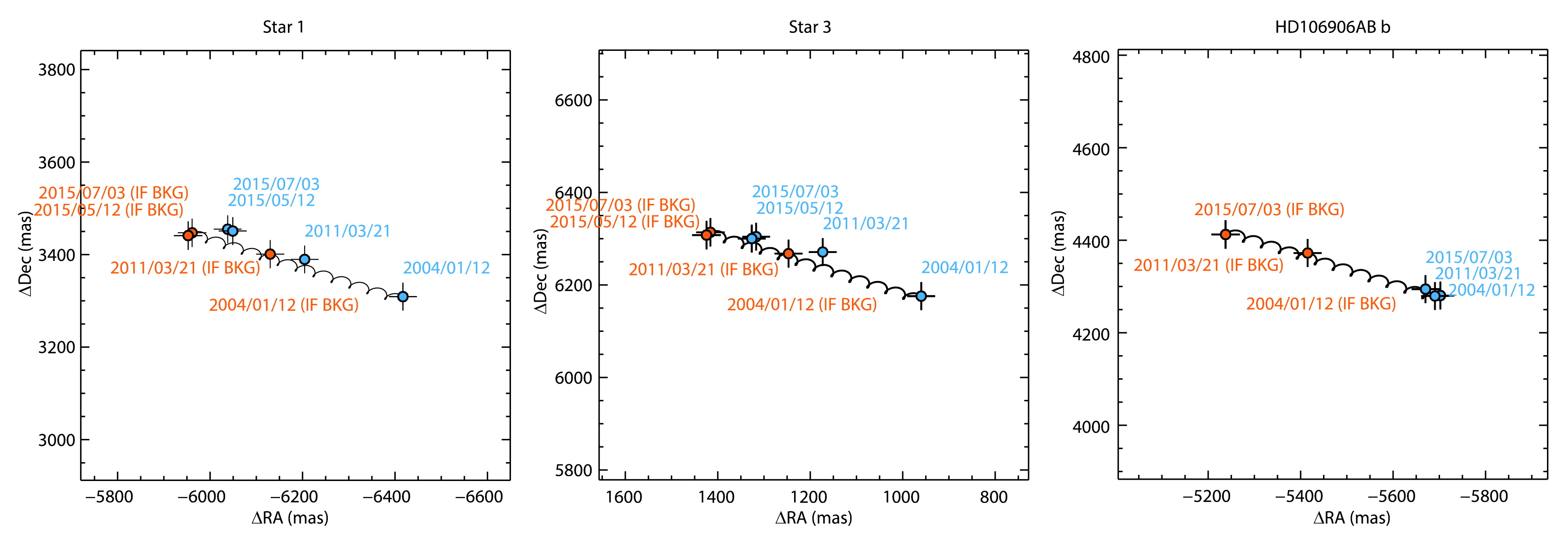}
     \caption{From left to right: { in blue, HD~106906AB b, star 1 and star 3 relative positions with respect to their positions at the first epoch (2004/01/12). The 2015 positions are given in this paper, and the other positions are taken from \cite{bailey14}}. The curve shows the expected motion if the targets are background objects. In red, the expected positions, assuming the targets are background targets, at each epoch. We adopted very conservative error bars (30 mas) for all data sets in this diagram, due to uncontrolled variations during the exposures.}
     \label{ppm}
\end{figure*}

Our IRDIS images reveal 3 additional point sources in the H2H3, H or K1K2 FoV (see Table\,\ref{cc}). Two of them (Star 1 and Star 2) were already present in ancillary HST ACS data (taken in 2004), and in the NICI planet b discovery images (taken in 2011). Given the long time span between the HST data (taken in 2004) and the present SPHERE data, and given HD~106906AB ppm, we could easily check that these sources do not share a common proper motion with HD~106906AB (see Fig.\ref{ppm}). Star 3\footnote{We label this target "Star" for convenience, but we do not have information yet on its nature.}, located closer to the star was not seen in NICI nor ACS data. Finally, another very faint target is seen in the H images south of Star 3, that was not present in the HST or NICI data. No point source was identified in the IFS FoV.

\begin{table}
\centering
\caption{Stars and planets around HD\,106906AB. PA and separations (in parenthesis, the month/day of observations, { all taken in 2015}) were measured using all images. The error bars provided here for the separation, PA, and the contrasts are those provided by the extraction procedure only. Additional photometric error bars  due to uncontrolled flux variations during the exposures are estimated to be 0.4 mag. Conservative astrometric error bars (dominated by uncontrolled variations plus distorsion effects) of 30 mas are estimated for the targets further than 6\arcsec \label{cc}}
\small
\begin{tabular}{@{}llll@{}}
 \hline
               & Sep  & PA    & $\Delta$Mag \\
               & mas    & deg   &     \\
\hline
Star 1         & 6965. $\pm$ 28. (07/03)       & 299.70 $\pm$ 0.23 (07/03)     & K1=12.7   $\pm$    0.1       \\ 
   &&&K2=12.6$\pm$0.2    \\
               &6957. $\pm$ 10. (05/12)&299.78 $\pm$ 0.1 (05/12)&H2=12.6    $\pm$ 0.1 \\
                &&&H3=12.5$\pm$    0.1\\
\hline
Star 2         &    6438.  $\pm$        20. (07/03) &  11.89    $\pm$     0.18 (07/03)  &  K1=12.0   $\pm$ 0.1   \\ 
               &      &     &    K2=11.8$\pm$ 0.2   \\
               &6440.  $\pm$  5. (05/12) & 11.80    $\pm$     0.05 (05/12)&H2=12.2   $\pm$ 0.1 \\
               &&&H3=12.0  $\pm$  0.1\\   
\hline
Star 3         &   2362. $\pm$ 16. (07/03)    &  68.99      $\pm$    0.40 (07/03)   &   K1=12.6 $\pm$   0.1 \\
               &     &  & K2=12.4 $\pm$  0.1\\  
               &2356. $\pm$ 4. (05/12) & 68.62    $\pm$     0.11 (05/12)&H2=12.6 $\pm$   0.1\\
               &&&H3=12.6  $\pm$  0.1 \\
\hline
Planet         &    7111.$\pm$13. (07/03)   &   307.15$\pm$0.1 (07/03)      &  K1=9.4   $\pm$ 0.1    \\
                  & &   &    K2=9.1$\pm$    0.1\\   
\hline
\end{tabular}
\end{table}

\begin{figure}[htp]
    \centering 
\includegraphics[trim = 12mm 5mm 4mm 5mm, clip,width=.4\textwidth]{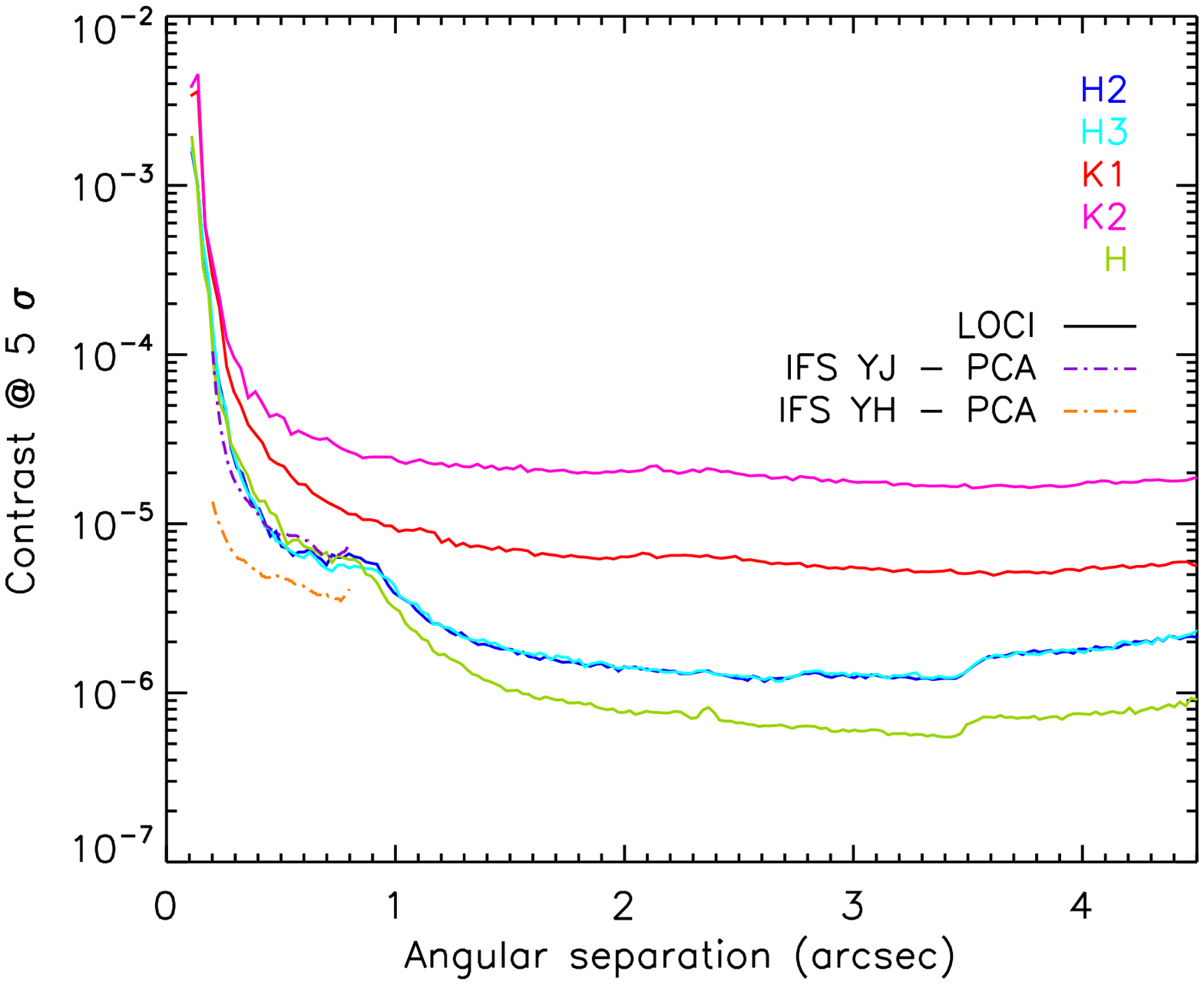}
\includegraphics[trim = 12mm 5mm 4mm 5mm, clip,width=.4\textwidth]{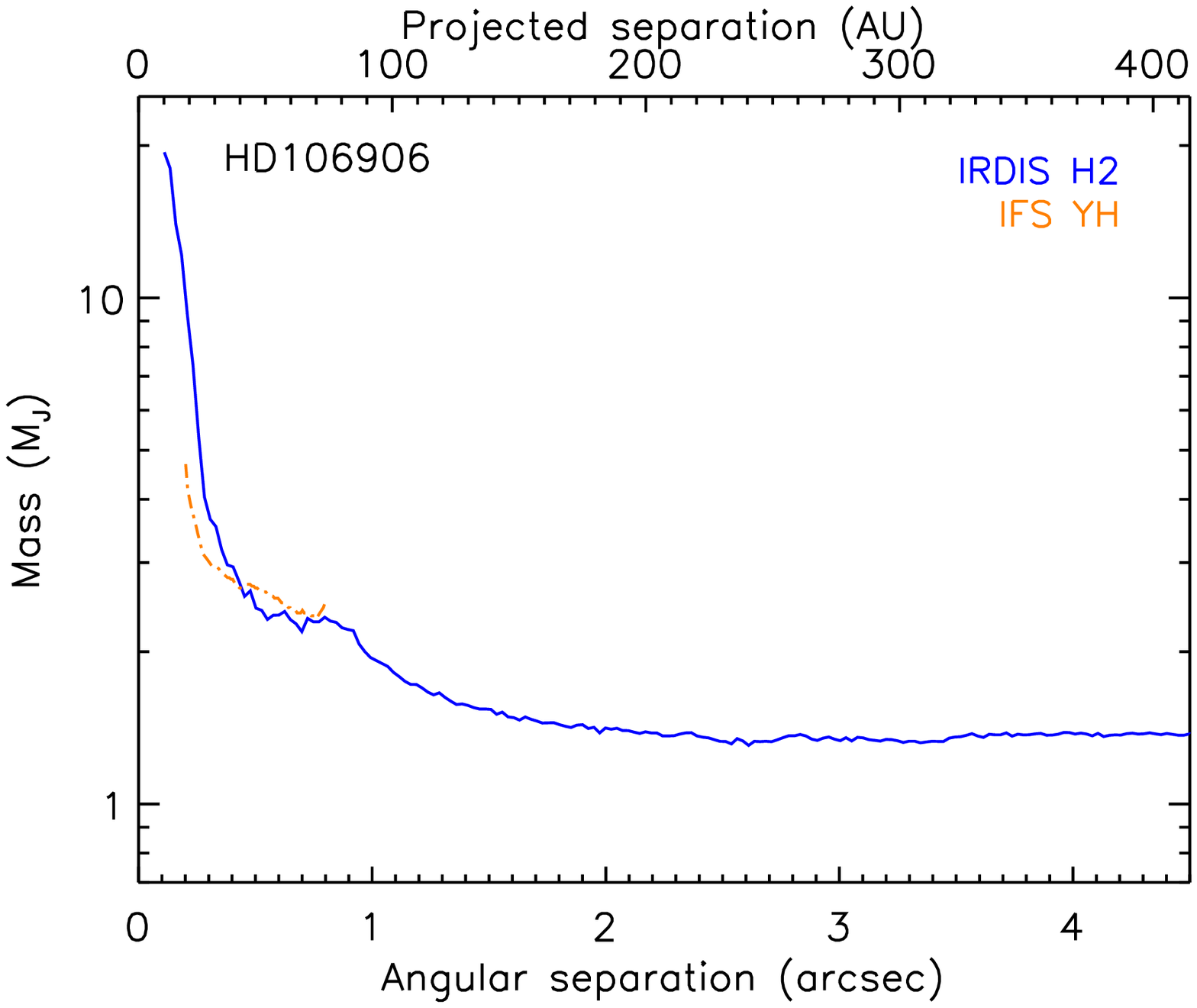}
\caption{Top: azimuthally averaged contrasts obtained with IRDIS  and with IFS. Bottom: best detection limits expressed in Jupiter masses. For IRDIS, the detection limits were combined to produce the best limits, once expressed in terms of masses. }
    \label{irdis_ifs_limdets}
  \end{figure}

Using the SPHERE IRDIS and IFS data, we computed the contrasts achieved in these observations. We then translated these contrasts into masses, using the BT-Settl + COND interior models adapted to the SPHERE filters, and assuming an age of 10 Myr. The results are summarized in Fig.\ref{irdis_ifs_limdets}. 
Note that the extinction towards HD~106906AB is negligible \cite[A$_{V}$=0.04 $\pm$ 0.02, ][]{pecaut12}.  We exclude companions with masses 1 \mjup ~or more at projected separations 200 au or more, planets with masses in the range 1-2 \mjup ~between 100 and 200 au, planets more massive than 3 \mjup within 30--100 au, and planets more massive than 10 \mjup ~in the range 10-30 au. These limits are significantly improved with respect to the ones (5-7 \mjup ~further than about 40 au) obtained with the Clio L' data \cite[][]{bailey14}. There is still the possibility that there are additional massive inner planets, possibly responsible for the inner void of material within the disk. They also leave room for additional planets that could be responsible for the inner edge, if orbiting closer than about 10 au from the inner edge, assuming that their separation (a), eccentricity (e), mass M$_{p}$ and distance to the edge $\delta a$ follow the \cite{mustill12} criterium : $\delta a$/a = 1.8 e$^{1/5}$ (M$_{p}$/M$_{*}$)$^{1/5}$.

\begin{acknowledgements}
The project is supported by CNRS, by the Agence Nationale de la Recherche (ANR-14-CE33-0018), and the Programme National de Plan\'etologie (PNP, INSU) and Programme National de Physique Stellaire (PNPS, INSU). A.L.M, D.M., and R.G. acknowledge support by Italian MIUR through "Premiale WOW 2013". JO acknowledges support from the Millennium Nucleus RC130007 (Chilean Ministry of Economy). We thank P. Delorme and E. Lagadec (SPHERE Data Center) for their efficient help during the data reduction process. SPHERE is an instrument designed and built by a consortium consisting of IPAG (Grenoble, France), MPIA (Heidelberg, Germany), LAM (Marseille, France), LESIA (Paris, France), Laboratoire Lagrange (Nice, France), INAF - Osservatorio di Padova (Italy), Observatoire de Genève (Switzerland), ETH Zurich (Switzerland), NOVA (Netherlands), ONERA (France) and ASTRON (Netherlands) in collaboration with ESO. SPHERE was funded by ESO, with additional contributions from CNRS (France), MPIA (Germany), INAF (Italy), FINES (Switzerland) and NOVA (Netherlands). SPHERE also received funding from the European Commission Sixth and Seventh Framework Programmes as part of the Optical Infrared Coordination Network for Astronomy (OPTICON) under grant number RII3-Ct-2004-001566 for FP6 (2004-2008), grant number 226604 for FP7 (2009-2012) and grant number 312430 for FP7 (2013-2016).  
\end{acknowledgements}
\bibliographystyle{aa}
\bibliography{biblio}

\end{document}